\documentclass[aps,pra,amsmath,amssymb,superscriptaddress,10pt,nobibnotes]{revtex4-2}
\usepackage{preamble}

\usepackage{siunitx}
\usepackage[justification=raggedright]{caption}
\usepackage{subcaption}
\usepackage{graphicx}
\usepackage{amsmath}
\usepackage{float}

\begin{document}


\title{Sensor fusion in ptychography.}

\author{Kira Maathuis}
\altaffiliation{\href{mailto:k.a.m.maathuis@uu.nl}{k.a.m.maathuis@uu.nl}}
\affiliation{Nanophotonics, Debye Institute for Nanomaterials Science and Centre for Extreme Matter and Emergent Phenomena, Utrecht University, P.O. Box 80000, 3508 TA Utrecht, The Netherlands}
\author{Jacob Seifert}
\affiliation{Nanophotonics, Debye Institute for Nanomaterials Science and Centre for Extreme Matter and Emergent Phenomena, Utrecht University, P.O. Box 80000, 3508 TA Utrecht, The Netherlands}
\author{Allard P. Mosk}
\affiliation{Nanophotonics, Debye Institute for Nanomaterials Science and Centre for Extreme Matter and Emergent Phenomena, Utrecht University, P.O. Box 80000, 3508 TA Utrecht, The Netherlands}


\begin{abstract}
Ptychography is a lensless, computational imaging method that utilises diffraction patterns to determine the amplitude and phase of an object. In transmission ptychography, the diffraction patterns are recorded by a detector positioned along the optical axis downstream of the object. The light scattered at the highest diffraction angle carries information about the finest structures of the object. We present a setup to simultaneously capture a signal near the optical axis and a signal scattered at high diffraction angles. Moreover, we present an algorithm based on a shifted angular spectrum method and automatic differentiation that utilises this recorded signal. By jointly reconstructing the object from the resulting low and high diffraction angle images, the resolution of the reconstructed image is improved remarkably. The effective numerical aperture of the compound sensor is determined by the maximum diffraction angle captured by the off axis sensor.

\end{abstract}

\maketitle


\section{Introduction}
Ptychography is a computational imaging method that relies on recording multiple diffraction patterns of an object while varying the illumination condition, often named the probe beam. These diffraction patterns are then  used to create a reconstructed image of the object. A thin object or 3D object\cite{Maiden:12,Godden:14} can be reconstructed, and the reconstruction consists of the amplitude and phase of the object \cite{Rodenburg:19}. The probe beam can be simultaneously recovered with the object \cite{Maiden:09}. The process of data acquisition introduces redundant information into the data, which ptychographic algorithms can use to accomplish compensation for experimental errors. This includes compensation for lateral scanning errors \cite{Hurst:10,Maiden:12B,Dwivedi:18}, to compensate axial positioning errors \cite{Dou:17,Lotgering:18,Lotgering:20}, and work with partially incoherent light sources \cite{Thibault:13}. 

Many algorithms exist to reconstruct ptychographical data, such as the ptychographical iteration engine (PIE) \cite{Rodenburg:04} family with ePIE \cite{Maiden:09}, mPIE  \cite{Maiden:17} and zPIE \cite{Lotgering:20}. Other notable algorithms are the difference map approach \cite{Thibault:08,Thibault:09} and the maximum likelihood (ML) approach \cite{Thibault:12}. Multiple different approaches can be combined and compared in a software framework PtyPy \cite{Enders:16}.

Recently, gradient-descent-based optimisation algorithms using automatic differentiation (AD) have been shown to be effective at solving the inverse problem in ptychography \cite{Kandel:19, Ghosh:18}. There is an essential distinction between conventional algorithms, such as the PIE family of algorithms, and algorithms using AD. The PIE algorithm iteratively applies an update function, and this update function needs to be derived in closed form for every experimental modification. For AD-based algorithms there is no need to derive such an update function manually. Instead, AD and an optimizer are employed to minimise the loss function that is calculated from the experimental data and the predicted diffraction patterns in the forward model of the system \cite{Ghosh:18}. As a result, automatic differentiation ptychography (ADP) is flexible to adaptations to the imaging system \cite{Kandel:19, Seifert:21} and can be used to work with complicated or difficult to invert systems.

The smallest resolvable object period, resolution $R$, of a ptychographic system is expressed as\cite{Claus:19},

\begin{equation} \label{eq:resolution}
    R = \lambda/(\textnormal{NA}_{\mbox{\scriptsize illum}} + \textnormal{NA}_{\mbox{\scriptsize  det}}), 
\end{equation}

\noindent where $\lambda$ is the illumination light's wavelength, $\textnormal{NA}_{\mbox{\scriptsize  illum}}$ numerical aperture of the illumination probe and $\textnormal{NA}_{\mbox{\scriptsize  det}}$ the numerical aperture of the detector. Enlarging the numerical aperture of the measured diffraction patterns offers a clear path to improved reconstruction resolution. Similarly, the use of structured illumination, such as speckle illumination, increases the NA of the incident light and enables higher resolution reconstructions \cite{Zhang:19,Odstrcil:19,Guizar-Sicairos:12}. Additionally, if the object of study can be characterised using a given set of parameters, we can optimise the illumination strategy using Fisher information \cite{Bouchet:21}.

\begin{figure}[htbp]
    \centering
    \begin{subfigure}[b]{1\columnwidth}
        \includegraphics[width=1\columnwidth]{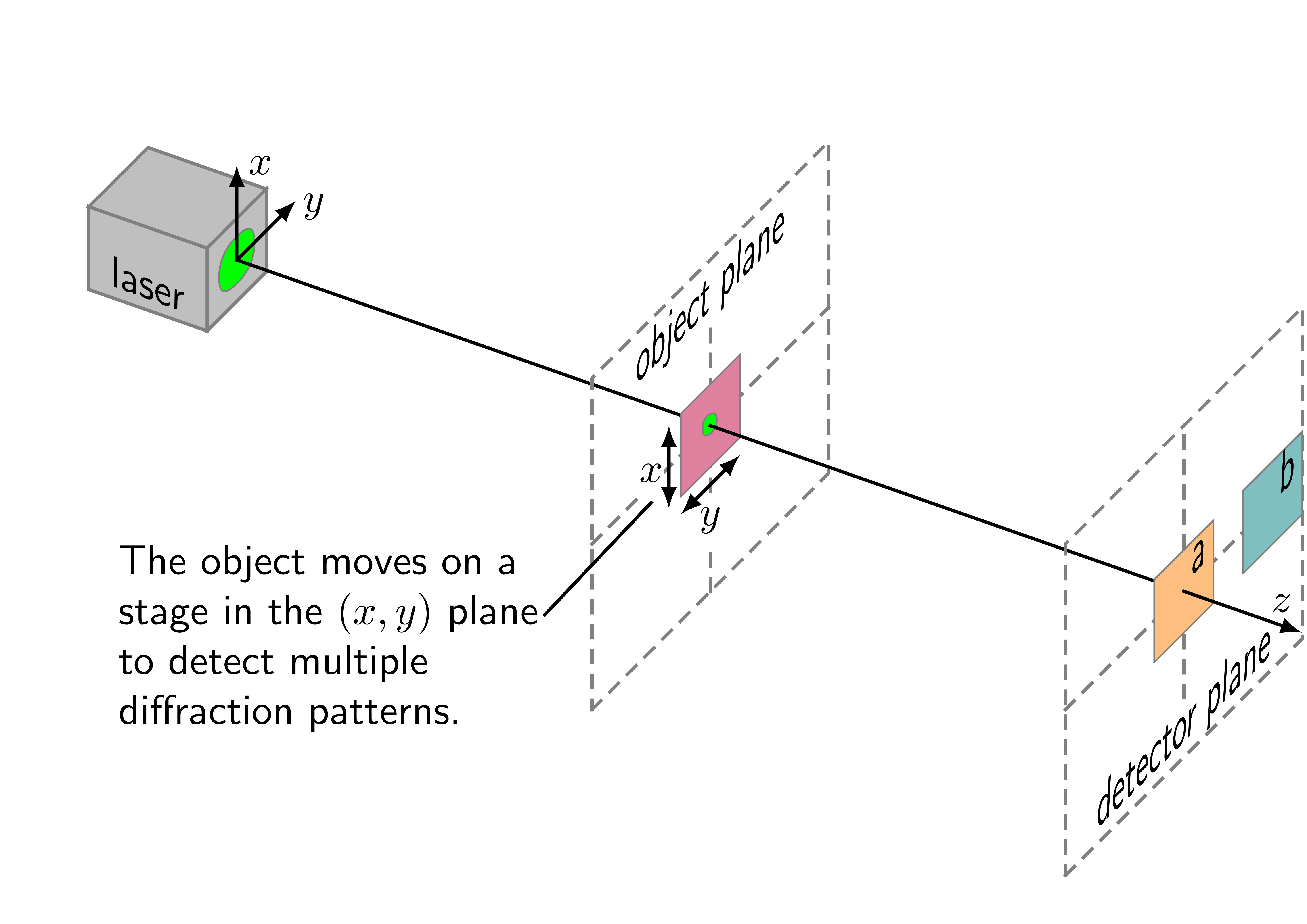}
    \caption{}
    \label{fig:setup}
    \end{subfigure}
    \vfill
    \begin{subfigure}[b]{0.49\columnwidth}
        \centering
        \includegraphics[width=\textwidth]{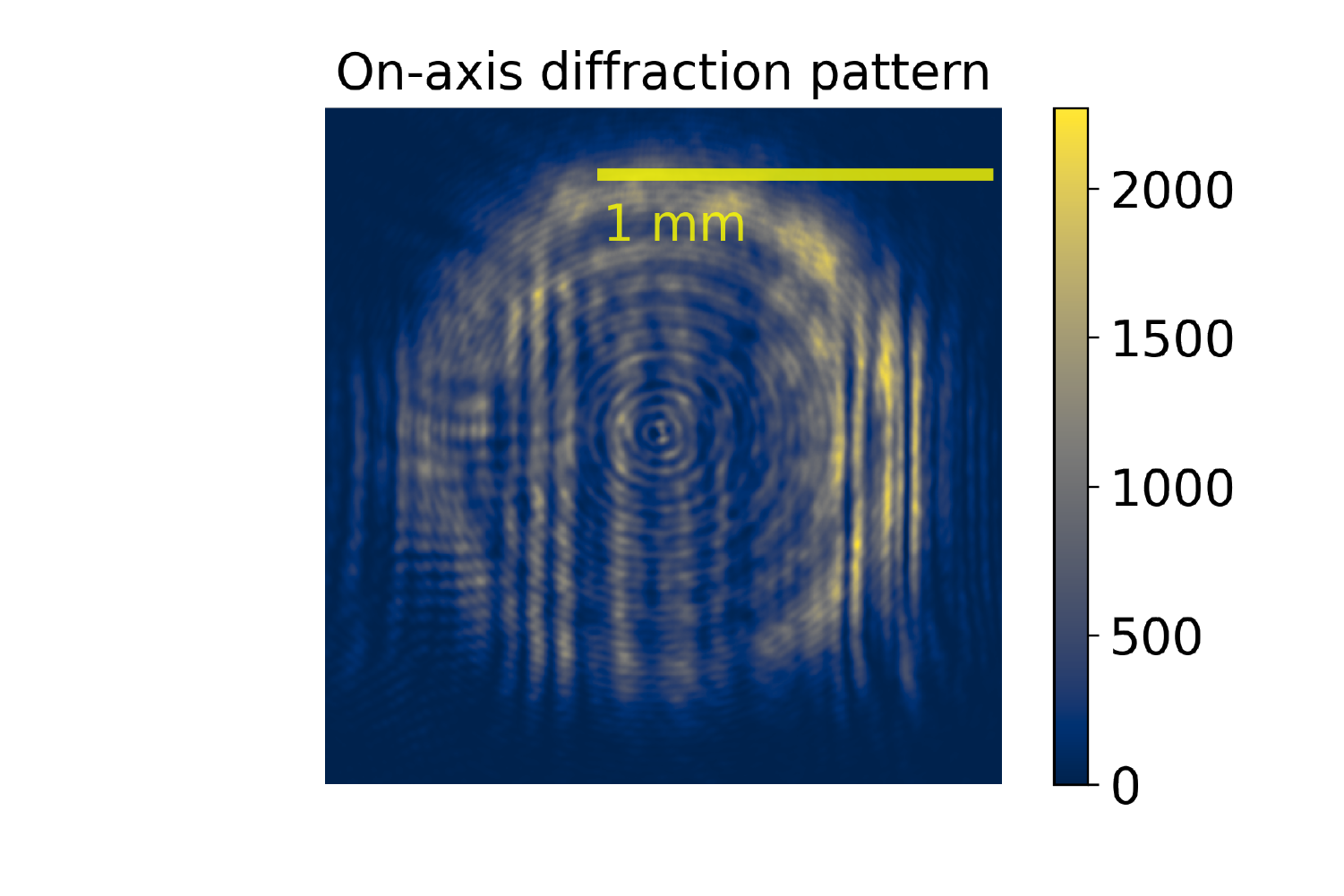}
        \caption{}
        \label{fig:onaxisdiffpat}
    \end{subfigure}
    \hfill
    \begin{subfigure}[b]{0.49\columnwidth}
        \centering
        \includegraphics[width=\textwidth]{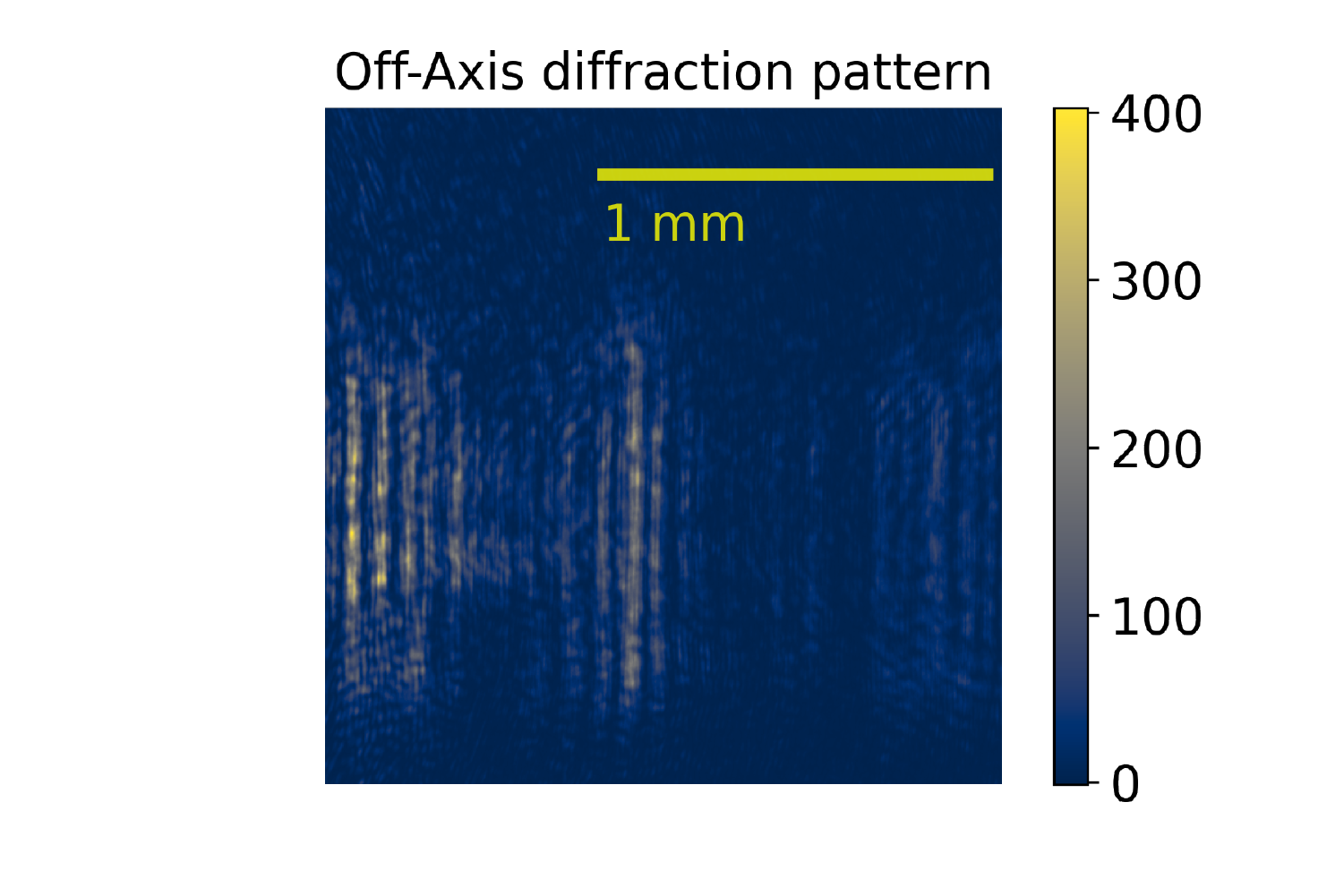}
        \caption{}
        \label{fig:offaxisdiffpat}
    \end{subfigure}
    \caption{\textbf{(a)} A diagram of the experimental setup. An aperture with a diameter of \SI{500}{\micro\metre} is illuminated with \SI{561}{\nano\metre} laser light and imaged to the object plane to form the probe field. The imaging optics and aperture are not shown in this figure. The probe field has a diameter of \SI{1.2}{\milli\metre} at the object. Together with the object it forms the exit field $\psi_{exit}(r,R_i)$ on the object plane. The object is laterally moved to scan positions $R_i$. At the detection plane, $\Psi_{det}(r',R_i)$, the orange region $a$ and teal region $b$ are recorded. \textbf{(b)} and \textbf{(c)} show experimental diffraction patterns of the regions of interest respectively. The colour bars represent the intensity.}
    \label{fig:setupandpatterns}
\end{figure}

In this paper, we introduce a ptychography system that includes a signal detected at high diffraction angles, hereafter coined the off-axis signal. This signal carries high spatial frequencies diffracted from the object, which is used to achieve an improved reconstruction resolution. We show our method of capturing the signal and method of reconstruction. 

\section{Problem formulation}
\label{sec:reconstruction model}

The additional signal is included using sensor fusion, by which we mean that data from different, physical or virtual, sensors is combined. In this work our physical detector is split into multiple sensor areas. The setup is shown in Fig.~\ref{fig:setup}. In the detector plane in Fig.~\ref{fig:setup}, two areas are distinguished representing the two sensors. One sensor measures the diffracted intensity close to the optical axis and the second one covers a disjunct area at higher diffraction angles.

We use the framework presented by Seifert et al.\cite{Seifert:21} to implement an algorithm that combines the on-axis and off-axis data. The algorithm consists of custom-built layers that together resemble the forward physics model of the experimental setup. The physics model is described as follows. At position vector $r$, the exit-field $\psi_\mathrm{exit}(r)$ in the object plane is formed by illuminating a complex-valued object $O(r)$ with a probe beam $P(r)$ at different positions $R_i$:

\begin{equation} \label{eq:exitfield}
    \psi_\mathrm{exit}(r,R_i)=O(r-R_i)P(r) \ .
\end{equation}

The exit-field is then propagated along the optical axis to the detector plane to form $\Psi_\mathrm{det}(r',R_i)$. The wave field propagation to the detector plane is calculated using the angular spectrum method. See supplemental document section 1 and accompanying code \cite{Maathuis:22} for exact implementation. The shifted angular spectrum method\cite{Matsushima:10} is implemented to propagate the wave field to an area of interest with a lateral offset to the optical axis. The intensity distribution $I_\mathrm{det}$ at the detector plane is given by

\begin{equation} \label{eq:detectorfield}
    I_\mathrm{det}(r',R_i)=|\Psi_\mathrm{det}(r',R_i)|^2 \ .
\end{equation}

\begin{figure}[htbp]
    \centering
    \includegraphics[width=1\columnwidth]{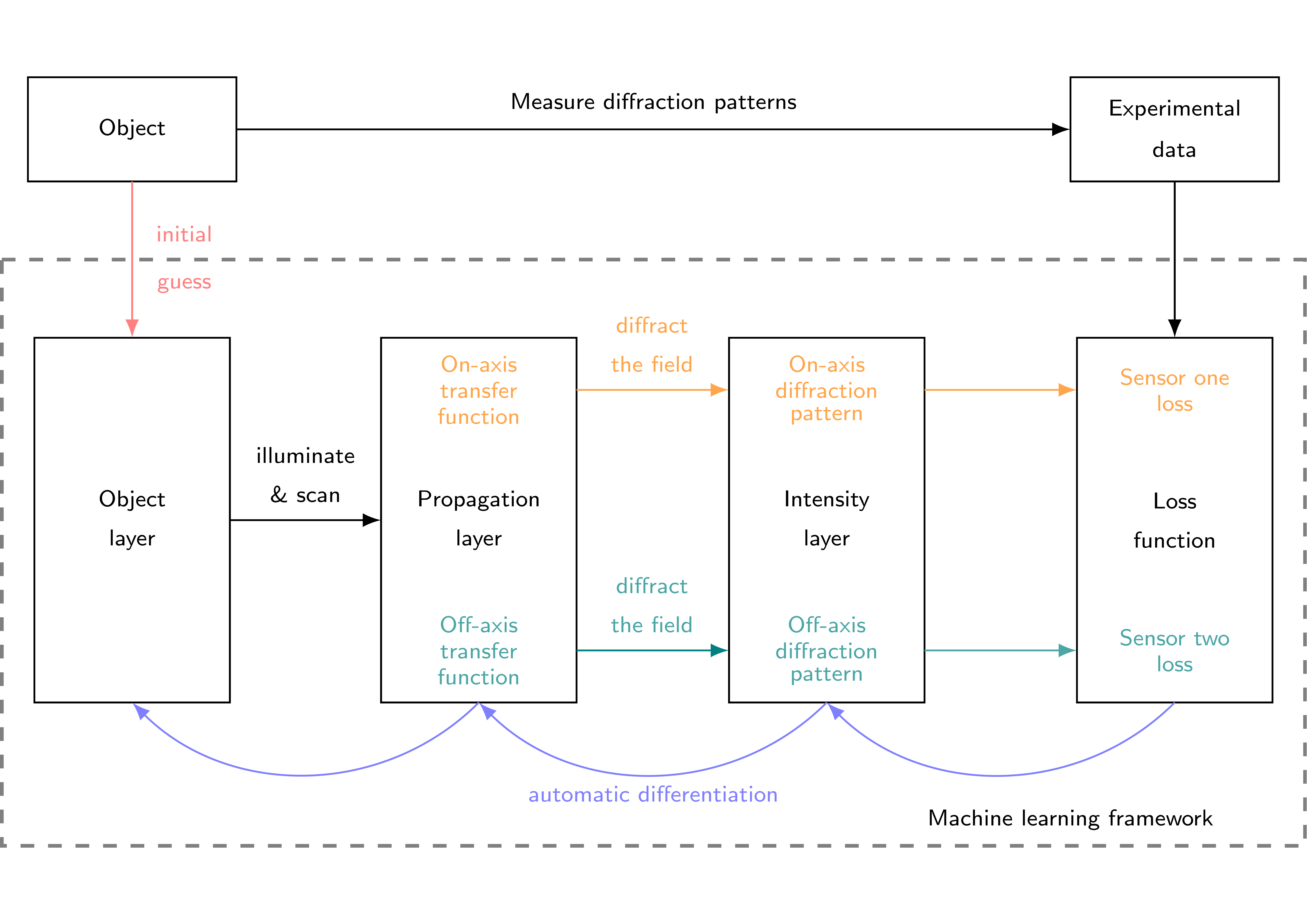}
    \caption{A flowchart of the ADP algorithm presented in this work adapted from the flowchart shown by Seifert et al. \cite{Seifert:21}, expanded to include the off-axis signal. The path of the data from left to right is the forward physics model. The on-axis data path is shown in orange, and the additional data path of the off-axis signal is shown in teal. The object layer starts with an initial guess, which is adapted by the automatic differentiation optimiser to minimise the MSE between the predicted and experimental diffraction patterns.}
    \label{fig:ADPmodel}
\end{figure}

A flowchart of the physics model and reconstruction routine is shown in Fig. \ref{fig:ADPmodel}. The object $O(r)$ and probe field $P(r)$ described in equation \eqref{eq:exitfield} compose the first layer. After the first layer the flowchart splits and the data flows diverge. The second layer is where the wave field in the object plane is diffracted to form the wave field in the detector plane by using the angular spectrum method. Here, the data is split into a path along the optical axis and a path away from the optical axis. Apart from selecting the correct diffraction model, this layer remains constant during operation. The last layer of the physics model converts the wave field at the detector plane to its intensity distribution using equation \eqref{eq:detectorfield}. At this point, there are two intensity distributions. One representing sensor $a$ and one representing sensor $b$. This constitutes the model prediction of the diffraction pattern $I_\mathrm{pred}(r,R_i)$ at a specific scanning position $R_i$. The cost function to be minimized is the mean squared error of the model prediction,

\begin{equation} \label{eq:MSE}
    \mathrm{MSE}=\frac{1}{N}\displaystyle\sum_{i=1}^{N}(I_\mathrm{meas}(r,R_i)-I_\mathrm{pred}(r,R_i))^2 \ ,
\end{equation}
with $N$ being the total number of points of the scanning pattern and $I_\mathrm{meas}(r,R_i)$ the experimentally measured diffraction pattern at position $R_i$. The MSE is computed for both the signal of sensor $a$ and the signal of sensor $b$. The relative weight of the two different MSE values is controlled by a mixing factor $\gamma$, which results in the final loss function $L$ given by 

\begin{equation} \label{eq:mixedMSE}
    L =(1-\gamma)\cdot \mathrm{MSE}_\mathrm{on-axis}+\gamma \cdot \mathrm{MSE}_\mathrm{off-axis}\ .
\end{equation}

\noindent There is an essential distinction between ptychography reconstruction based on conventional algorithms, such as PIE, and algorithms using AD. In comparison, PIE uses a closed-form update function; in ADP, the object retrieval results from minimising a loss function using the reverse mode of AD. In this model, $L$ is minimised using multiple iterations of the Adam optimiser \cite{Kingma:15}, which implements automatic differentiation. Consequently, this algorithm leads to the reconstruction of the object transmission function stored in the first layer of the model.

ADP uses efficient optimisers developed by the machine learning communities. The algorithm is implemented using the Tensorflow\cite{Abadi:16} and Keras\cite{Chollet:15} libraries. The Keras library allows us to map the reconstruction task in ptychography onto an architecture similar to deep or multi-layered neural networks. The physics-based layer-by-layer approach makes the algorithm highly modular, as layers can be modified or new layers added to extend the physics model. Additionally, to control the optimisation process, the ability to use alternative loss functions is paramount, such as incorporating two data flows as in equation \eqref{eq:mixedMSE}.

\section{Experimental setup}

The probe field $P(r)$ is formed by shining a laser of wavelength \SI{561}{\nano\metre} (Cobolt Jive 100) at an aperture with a diameter of \SI{500}{\micro\metre}. The aperture is imaged to the object plane to form the probe field. The probe field has a diameter of \SI{1.2}{\milli\metre} at the object plane. The object is mounted on a stage (stepper motor Thorlabs ZFS25B, controller Thorlabs KST101) that is laterally moved to a total of 52 positions $R$ with an overlap of $85\%$, above the overlap requirement shown by Bunk et al.\cite{Bunk:08}. Section 2 in the supplemental material describes the scanning pattern. Our test object, a Thorlabs NBS 1952 resolution target, consists of sets of three horizontal and vertical lines. Numbers indicate the lines per \si{\milli\metre} of each set. The light that is diffracted by the object  is lenslessly captured on a camera chip (Basler ace acA2440-35um) placed \SI{61}{\milli\metre} behind the object. We define two virtual sensors on the camera chip as shown in Fig.~\ref{fig:setup}.  Sensor \textit{a} captures the diffraction pattern propagating close to the optical axis and sensor \textit{b} captures a part of the diffraction pattern shifted \SI{2649}{\micro\metre} away from the optical axis. The effect of shifted distance and sensor size on reconstruction quality is discussed in section 3 of the supplemental material.

In Fig. \ref{fig:onaxisdiffpat} and Fig. \ref{fig:offaxisdiffpat} we show examples of diffraction patterns captured by sensors \textit{a} and \textit{b} respectively. Each virtual sensor region contains $512\times512$ pixels. Sensor $b$ only captures light that is diffracted at large horizontal angles. The $\textnormal{NA}$ of the system is not changed isotropically. While the $\textnormal{NA}$ is increased in the horizontal direction, the $\textnormal{NA}$ of the system remains the same in the vertical direction. As equation \eqref{eq:resolution} shows, the resolution of the system depends on its $\textnormal{NA}$, we expect the extra signal captured by sensor $b$ not to contribute isotropically to a higher resolution and instead only to a higher horizontal resolution.For each of the 52 scanning positions, two images were recorded by the detector. To properly capture the off-axis signal, the exposure time for  sensor \textit{b} was increased to 10 times the exposure time of  sensor \textit{a}. One image was taken using an exposure time of \SI{50}{\milli\second} for sensor $a$ and one using an exposure time of \SI{500}{\milli\second} sensor $b$.


\section{Results}

\begin{figure}[htbp]
        \centering
        \includegraphics[width=1\columnwidth]{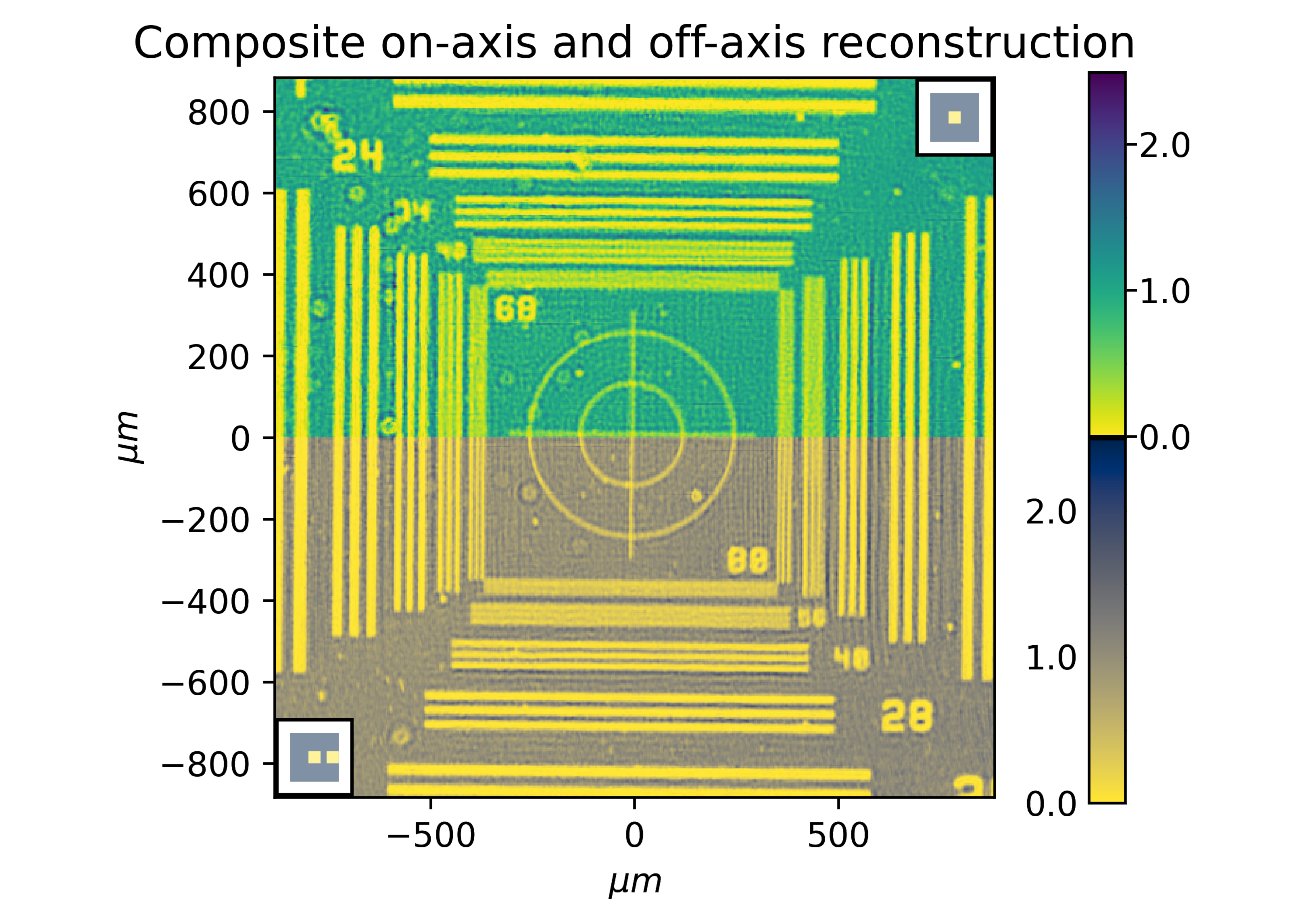}
        \caption{A composite image of two reconstructions. The upper half of the image represents the transmittance of an object reconstruction from experimental data using only sensor $a$ along the optical axis. The lower half of the image, represents the transmittance using sensor $b$ additionally to sensor $a$. The active sensor areas are indicated in the inset images in the bottom left and top right corner. The colour bars represent the transmittance of the reconstructed object. The reconstructed transmittance was normalised to 1 for an area on the target known to be clear glass.}
        \label{fig:transmittance}
\end{figure}

In Fig. \ref{fig:transmittance}, we show a composite image of two individual reconstructions. In the upper half, the object reconstructed only from diffraction patterns of sensor $a$ is shown.
In the lower half, both sensors are utilised to reconstruct the object.
Both images are retrieved using an Nvidia RTX 2070 GPU by running our reconstruction algorithm for 40\,epochs taking approximately 120 seconds. The mixing factor $\gamma$ in equation \eqref{eq:mixedMSE} is set to $0$ for the reconstruction in the upper half of Fig.~\ref{fig:transmittance}. For the reconstruction in the lower half, $\gamma$ is switched from $0$ to $0.5$ after 20 epochs. A mixing factor of $0.5$ means that from that point onward, the MSE of the on-axis signal and off-axis signal are both taken into account with similar weight. The measured diffraction patterns are normalized to the exposure time.

We observe for the sets of vertical lines that the highest density of lines that is resolved in the upper half of the image is 48\,lines/mm, i.e. a spacing of \SI{20.8}{\micro\metre}. Any denser set of lines remains unresolved for the on-axis reconstruction. In the reconstruction that incorporates the off-axis signal, 80\,lines/mm are resolved, equivalent to \SI{12.5}{\micro\metre}. The 80\,lines/mm set is the finest set of lines present in our target. Therefore, in the particular geometry of our experimental setup, we gain an improvement in resolution of at least a factor 1.6. Note that, as expected, we do not observe an improved vertical resolution. 

\begin{figure}[htbp]
        \centering
        \includegraphics[width=1\columnwidth]{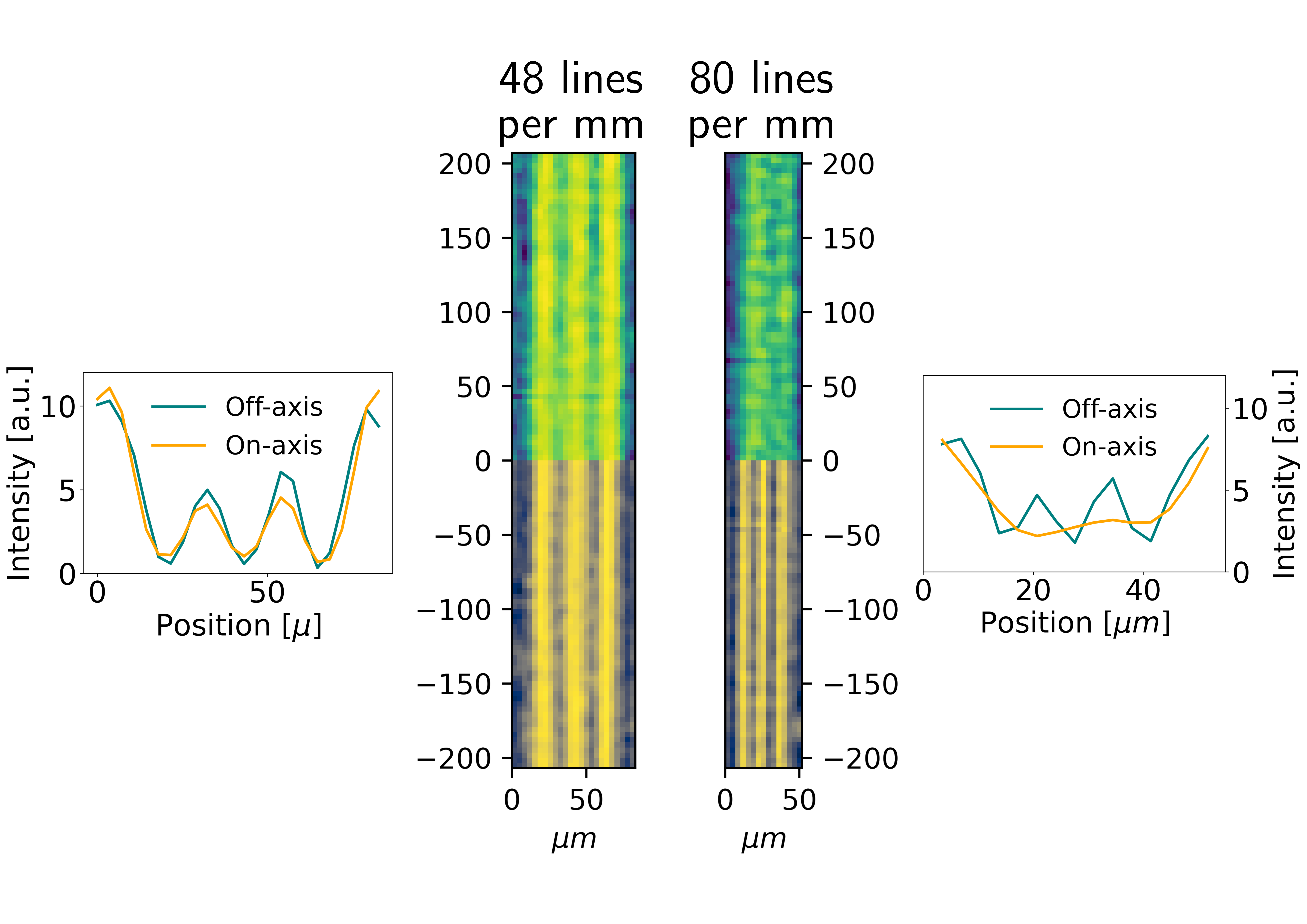}
        \caption{A cropped view of the vertical sets of 48 and 80 lines per mm from Fig.~\ref{fig:transmittance}. The upper half of each image shows the on-axis reconstruction, and the lower half the mixed on-axis and off-axis reconstruction. The red intensity curves depict the averaged values along the full vertical direction in their respective.}
        \label{fig:48and80linescomparison}
\end{figure}

In Fig.~\ref{fig:48and80linescomparison}, a cropped view of two sets of vertically oriented lines is shown, the 48\,lines/\si{\milli\metre} and the 80\,lines/\si{\milli\metre} set. The upper and lower half are the same reconstruction images as shown in Fig.~\ref{fig:transmittance}. Aside of the reconstructed image, lines show the intensity values as if we look at one row of pixels along the horizontal axis. These values are averaged values from each row along the full length of the lines. In this image, low intensity represents a line where the light is blocked, and high intensity the surrounding area where the light passes. In Fig.~\ref{fig:48and80linescomparison} at 48\,lines/\si{\milli\metre}, we see a pattern of three $I_{min}$ values and two $I_{max}$ values, which is the three-lined pattern of each line set. The pattern is visible for both reconstructions at 48 lines per mm. However, for 80 lines per \si{\milli\metre}, this pattern is only visible for the combined on-axis and off-axis signal reconstruction. The line pattern is not resolved for the on-axis signal reconstruction.

The fringe visibility ($V$) \cite{Ellis:14} is a measure to quantify the resolution of these line patterns, and is given by 

\begin{equation}
    V = \frac{I_\mathrm{max}-I_\mathrm{min}}{I_\mathrm{max}+I_\mathrm{min}} \ .
    \label{eq:fringecontrast}
\end{equation}

\noindent Here, $I_\mathrm{max}$ and $I_\mathrm{min}$ denote the maximum and minimum intensities of the signal, respectively. The calculation of $V$ is performed using the mean values of the two bright $I_{max}$ values between the three $I_{min}$ lines and the mean value of these three $I_{min}$ values. All mean values are calculated along the full lengths of the lines.

Table~\ref{tab:fringecontrast} shows several fringe visibility values computed for the transmittance reconstructions shown in Fig.~\ref{fig:transmittance} and Fig.~\ref{fig:48and80linescomparison}. Empty values indicate that the line set is not clearly resolved, which occurs when $I_\mathrm{max}$ and $I_\mathrm{min}$ are not distinguishable. For the lines with horizontal orientation, we observe no significant difference in resolution between the on- and off-axis object reconstructions. Both reconstructions have a decent fringe visibility for the set 48\,lines/\si{\milli\metre}, and a poor fringe visibility for the set of 56\,lines/\si{\milli\metre}.

Analysing the fringe visibility for the line sets with vertical orientation, we observe a significant improvement in fringe visibility for the object reconstruction that incorporates the off-axis signal. The fringe visibility of this reconstruction at 80\,lines/\si{\milli\metre} is similar to the fringe visibility of the on-axis signal reconstruction at 48\,lines/\si{\milli\metre}.

\begin{table}[htbp]
    \centering
    \caption{\bf Fringe visibility of the line sets within the transmittance reconstructions. Empty values indicate that the line set is not clearly resolved which occurs when $I_\mathrm{min}$ and $I_\mathrm{max}$ are not distinguishable.}
    \begin{tabular}{cccc}
        \hline
        lines per mm & line set & on-axis fringe & mixed signal fringe \\
         & orientation & visibility & visibility \\
        \hline
        48 & vertical & 0.61 & 0.86 \\
        56 & vertical & - & 0.68 \\
        80 & vertical & - & 0.55 \\
        \hline
        48 & horizontal & 0.52 & 0.48 \\
        56 & horizontal & - & - \\
        \hline
    \end{tabular}
    \label{tab:fringecontrast}
\end{table}


\section{Discussion}
The reconstruction using only on-axis data, with $\textnormal{NA}_{\mbox{\scriptsize illum}} = 0.01\pm 0.003$, and $\textnormal{NA}_{\mbox{\scriptsize det}} = 0.015$ has a theoretical resolution  between \SI{20.0}{\micro \metre} and \SI{25.5}{\micro \metre} according to Eq.~\eqref{eq:resolution}. Using the off axis data as well increases the detection NA to $\textnormal{NA}_{\mbox{\scriptsize det}} = 0.036$ leading to  a resolution  between \SI{11.4}{\micro \metre} and \SI{13.0}{\micro \metre} in the horizontal direction. The off axis data does not change vertical NA, so no effect on the vertical resolution is expected.

Both our reconstruction based on solely on-axis data and our reconstruction based on the fused on-axis and off-axis data show a resolution that corresponds well to the initially estimated resolution. For the on-axis case the best measured resolution is \SI{20.8}{\micro\metre} and for the fused data we find a resolution of \SI{12.5}{\micro\metre}. This is within the margin of error with which we estimate the illumination numerical aperture.
 
Remarkably, the gain in resolution does not require a sensor that spans the entire solid angle. In fact the off-axis sensor samples only light reflected to the right of the optical axis. For samples with a pronounced 3D structure, for example a blazed transmission grating, this may lead to incorrect reconstructions and a symmetric sensor configuration may be preferable. However if the object is known to be two-dimensional it is sufficient to capture only angles on one side, comparable to the use of single-sideband methods in signal processing \cite{Haykin:78}.

The presented results demonstrate the strength of optimization-based approaches to ptychography, allowing it to be extended to incorporate sensors at different positions and orientations, and in principle even of entirely different types. If the relative position of different sensors is unknown, an intriguing possibility is to use the inherent redundancy in ptychographic data to correct such detectors' positions, similar to the axial and lateral corrections previously shown. 

\newpage


\section{Conclusion}
To conclude, we show that the fusion of two laterally displaced sensors in the detector plane improves the resolution in ptychographic reconstructions. We achieve this improvement by extending an optimization-based ptychography framework with two separate data branches to calculate the diffraction intensity distributions in two spatially separated sensors in the detection plane, of which one is placed off the optical axis. By tuning the weight of the individual sensors in the optimization cost fucntion, we can then choose the relative contribution of each sensor to the reconstruction process. Experimentally, sensor fusion results in a a significant improvement in the resolution of a reconstructed target object, which agrees with a resolution estimated based on the maximum diffraction angle captured by the off-axis sensor.

By fusing multiple sensors, our work enables new applications of ptychography where several small sensors, possibly with a tuned sensitivity and gain, cover a wide numerical aperture in a way that is more flexible than can be achieved with a single sensor. For instance a robust low-sensitivity sensor could be used to detect the high on-axis intensities while sensitive high-gain sensors are used for the weak diffraction patterns at large angles. \\
\linebreak
\noindent \textbf{Funding.} Netherlands Organization for Scientific Research NWO \\ (Vici 68047618 and Perspective P16-08)

\noindent \textbf{Disclosures.} The authors declare no conflict of interest.

\noindent \textbf{Data Availability.} Data and analysis methods supporting this work are available in Ref. \cite{Maathuis:22}.

\newpage

\section*{Sensor fusion in ptychography: supplemental document}

This document provides supplementary material to \textit{Sensor fusion in ptychography}.

\subsection{Implementation of propagation between parallel planes.}

\begin{figure}[H]
    \centering
    \begin{subfigure}[b]{0.49\columnwidth}
        \centering
        \includegraphics[width=\textwidth]{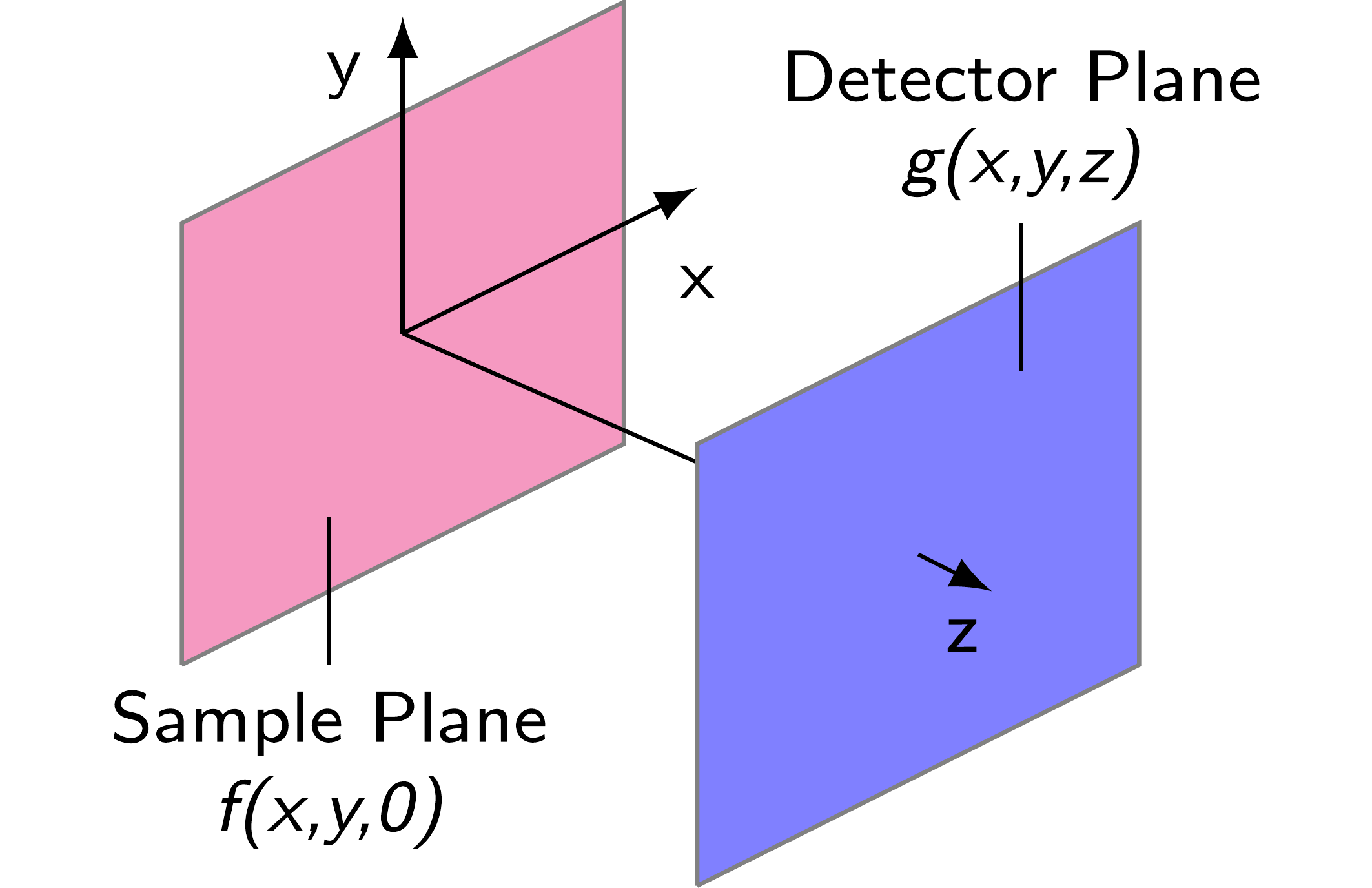}
        \caption{}
        \label{fig:ASPW}
    \end{subfigure}
    \hfill
    \begin{subfigure}[b]{0.49\columnwidth}
        \centering
        \includegraphics[width=\textwidth]{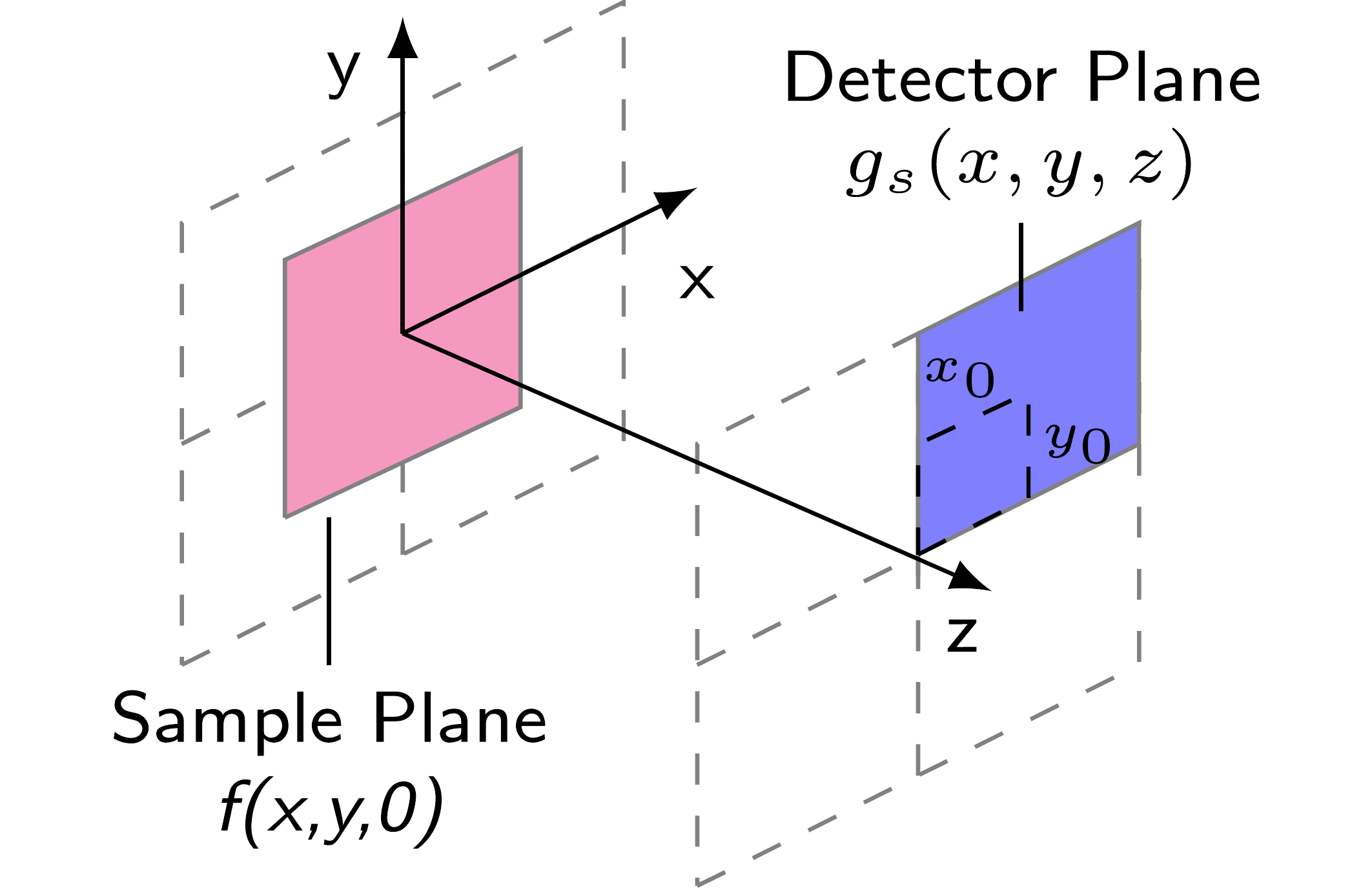}
        \caption{}
        \label{fig:shiftASPW}
    \end{subfigure}
    \caption{ \textbf{(a)} The geometry and coordinate system of the parallel plane model for the angular spectrum. \textbf{(b)} The geometry and coordinate system of shifted plane model for the angular spectrum. Note that the centre of the detector plane is shifted by $x_0$ and $y_0$ distance from the central z-axis}\label{fig:ASprop}
\end{figure}

The propagation of a field from a sample plane to parallel detection plane a distance $z$ away, as seen in figure \ref{fig:ASPW}, with the angular spectrum method takes three steps. Given the field $f(x,y,0)$ on the sample plane, step one is transforming to the frequency domain by Fourier transformation

\begin{equation}
  F(f_x,f_y;0) \! = \! \iint\limits_\infty^\infty \! f(x,y,0)e^{-i2\pi(f_xx+f_yy)}dxdy \qquad .
  \label{FTsample}
\end{equation}

The frequency-domain field $F(f_x,f_y;0)$ is then multiplied with the propagator $H$, also known as the \textit{transfer function}, to propagate the field towards $z$ \cite{Goodman:96}. This will give the frequency-domain field $G(f_x,f_y;z)$ on the detector plane,

\begin{equation*}
  G(f_x,f_y;z) \! = \! F(f_x,f_y;0)H(f_x,f_y;z) \ ,
\end{equation*}
\begin{equation}
  H(f_x,f_y;z) \! = \!e^{i\frac{2\pi}{\lambda}z\sqrt{1-(\lambda f_x)^2-(\lambda f_y)^2}} \qquad .
  \label{eq:prop}
\end{equation}

The previous step yielded the field at $z$ in the frequency domain and to find the spatial-domain field at the detector plane, $g(x,y,z)$, we only have to perform one last step, compute the inverse Fourier transform

\begin{equation}
  g(x,y,z) \! = \! \iint\limits_\infty^\infty \! G(f_x,f_y;z)e^{i2\pi(f_xx+f_yy)}dxdy \ .
  \label{IFTdetector}
\end{equation}

The implementation of the angular spectrum method just described is limited in its scope by the fact that the sample plane and detector plane share the same optical axis and their centre lies upon that axis. For fields that do not travel along that optical axis or for diffracted light that falls far off this optical axis, it is useful to be able to compute the field found at a detector plane laterally shifted from the axis. We can compute this by simply expanding the array used to represent the field to include the area of interest and cutting out this area. However, this is computationally inefficient, as most of the array will be of little interest and the guard bands will need to be enormous.

This means we need a method to compute off-axis numerical propagation, and this is done by modifying the angular spectrum method. Following is a shortened version of the modification of the angular spectrum method presented by K. Matsushima \cite{Matsushima:10}.

Fig. \ref{fig:shiftASPW} shows the geometry of this situation. The detector plane $g_s(x,y,z)$ is shifted away from the optical axis (z-axis) by a distance of $x_0$ and $y_0$. The maximum and minimum spatial frequencies falling on the detector plane can be expressed as $u_{max}$, $u_{min}$, $v_{max}$ and $v_{min}$, where $u$ represents the spatial frequencies for the $x$ direction and $v$ for the $y$ respectively. The formulae to compute these are shown below in equation \eqref{eq:UVmaxUVmin}, with $\lambda$ being the light's wavelength,

\begin{equation*}
  u_{max} = sin(\theta_{max})/\lambda \ ,
\end{equation*}
\begin{equation*}
  u_{min} = sin(\theta_{min})/\lambda \ ,
\end{equation*}
\begin{equation*}
  v_{max} = sin(\theta_{max})/\lambda \ ,
\end{equation*}
\begin{equation}
  v_{min} = sin(\theta_{max})/\lambda \ .
  \label{eq:UVmaxUVmin}
\end{equation}

Angles $\theta_{min}$ and $\theta_{max}$ represent the minimum and maximum spatial frequencies that can still reach the area of interest shown in figure \ref{fig:shiftASPW}. The computation of these angles is useful for determining the spatial frequencies present on the detector plane and their determination is paramount for the proper integration of a shifted angular spectrum method. They will be used to cut the spatial frequencies that do not reach the detector in the frequency domain.

However, before we cut the incorrect spatial frequencies, the transfer function $H(f_x,f_y;z)$ from equation \eqref{eq:prop} needs to be modified to accommodate the shift of $x_0$ and $y_0$. In equations \eqref{eq:shiftFT} and \eqref{eq:shiftprop}, $G_s(f_x,f_y;z)$ represents the detector plane in the frequency domain and $H_s(f_x,f_y;z)$ the modified transfer function. Note that the transfer function consists of the original transfer function and an addition that incorporates $x_0$ and $y_0$

\begin{equation}
  Gs(f_x,f_y;z) \! = \! F(f_x,f_y;0)H_s(f_x,f_y;z) \ ,
  \label{eq:shiftFT}
\end{equation}
\begin{equation}
  H_s(f_x,f_y;z) \! = H(f_x,f_y;z)\:e^{i2\pi(x_0f_x+y_0f_y)} \ .
  \label{eq:shiftprop}
\end{equation}

Using this modified transfer function, the shift to the new detector plane is achieved. To avoid sampling problems in the form of aliasing errors, only the spatial frequencies present on the detector plane must be propagated. As long as the distance $z$ is larger than $S_x$ and $S_y$ this can be approximated by a cut the frequency domain using a rectangular function \cite{Matsushima:10}. Using $u_{max}$, $u_{min}$, $v_{max}$ and $v_{min}$ we can determine a width ($u_{width}$ and $v_{width}$) and displacement ($u_0$ and $v_0$) for the rectangle cut. These variables are used to effectively band-limit the transfer function $H_s$ (also applicable to $H$). They are dependent on certain cases, as described in table \ref{tab:cases}. In equation \ref{eq:shift}, $rect$ represents the rectangular function

\begin{equation}
    \begin{split}
    Gs(f_x,f_y;z) \! = \! F(f_x,f_y;0)H_s(f_x,f_y;z) rect\left(\frac{f_x-u_0}{u_{width}}\right)rect\left(\frac{f_y-v_0}{v_{width}}\right) . \:
    \end{split}
    \label{eq:shift}
\end{equation}

Finding the field $g_s(x,y;z)$ once more becomes an inverse Fourier transformation of $G_s(f_x,f_y;z)$. With this supplementary document comes the \textsc{Python} source code of ADP, showing the implementation of this method.

\begin{table}[H]
  \centering
  \caption{\bf Cases and variables needed to band-limit the transfer function $H_s$ as shown in \cite{Matsushima:10}.}
  \small
  \begin{tabular}{|l|l|l|}
    \hline
    case & $u_0$/$v_0$ & $u_{width}$/$v_{width}$ \\
    \hline
    
    $S_x<x_0$ & $(u_{max}+u_{min})/2$ & $u_{max}-u_{min}$ \\
    $S_y<y_0$ & $(v_{max}+v_{min})/2$ & $v_{max}-v_{min}$ \\
    \hline
    
    $-S_x \leq x_0 < S_x$ & $(u_{max}-u_{min})/2$ & $u_{max}+u_{min}$ \\
    $-S_y \leq x_0 < S_y$ & $(v_{max}-v_{min})/2$ & $v_{max}+v_{min}$ \\
    \hline
    
    $x_0 \leq -S_x$ & $-(u_{max}+u_{min})/2$ & $u_{min}-u_{max}$ \\
    $y_0 \leq -S_y$ & $-(v_{max}+v_{min})/2$ & $v_{min}-v_{max}$ \\
    \hline
  \end{tabular}
  \label{tab:cases}

\end{table}

\subsection{Scanning trajectories}

\begin{figure}[H]
    \centering
    \includegraphics[width=0.8\textwidth]{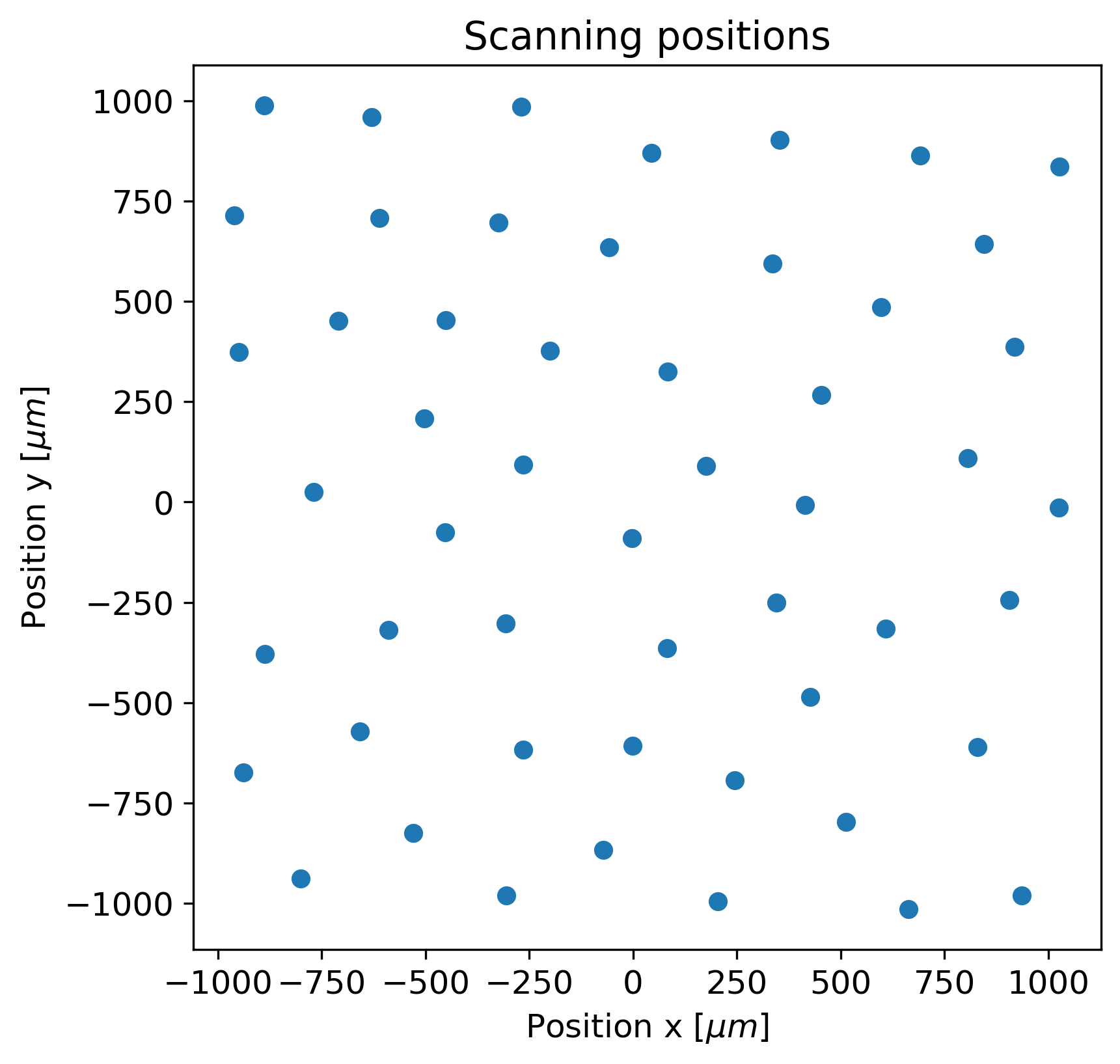}
    \caption{The 52 scanning positions $R_i$ used to gather the experimental data. The positions are Poisson disk sampling points generated using a minimum distance between scanning points of $r= \SI{0.25}{\milli\metre}$.}
    \label{fig:scanning_positions}
\end{figure}

Fig. \ref{fig:scanning_positions} shows the scanning positions used to record the experimental data. Raster grid scanning of the experimental sample was avoided by using a scanning pattern based on Poisson disk sampling  \cite{Bridson:07}. In total, 52 scanning positions were generated with a minimum distance between scanning points of $r= \SI{0.25}{\milli\metre}$.

Poisson disk sampling is a sampling scheme that has a non-periodical and uniform distribution, which has been shown to be beneficial to ptychography \cite{Huang:14}.

\subsection{Full frame and other sensor configuration reconstructions}

\begin{figure}[H]
    \centering
    \begin{subfigure}[b]{0.49\columnwidth}
        \centering
        \includegraphics[width=\textwidth]{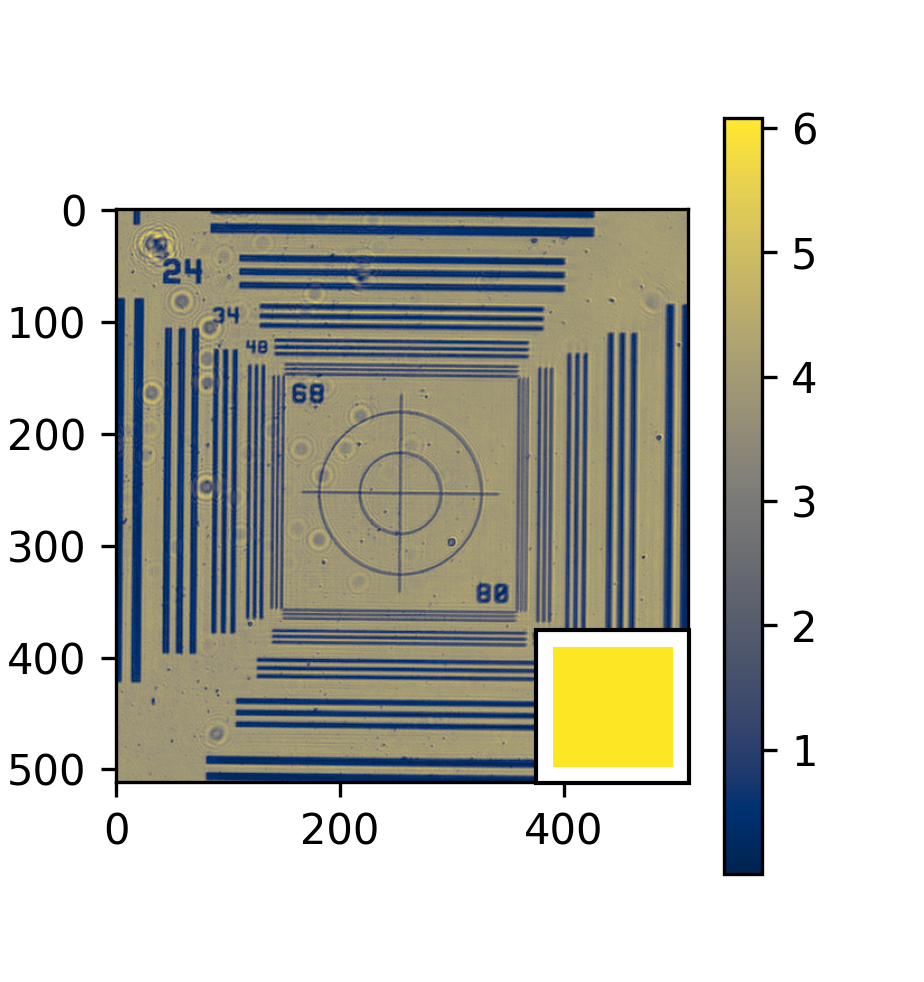}
        \caption{}
        \label{fig:fullframe}
    \end{subfigure}
    \hfill
    \begin{subfigure}[b]{0.49\columnwidth}
        \centering
        \includegraphics[width=\textwidth]{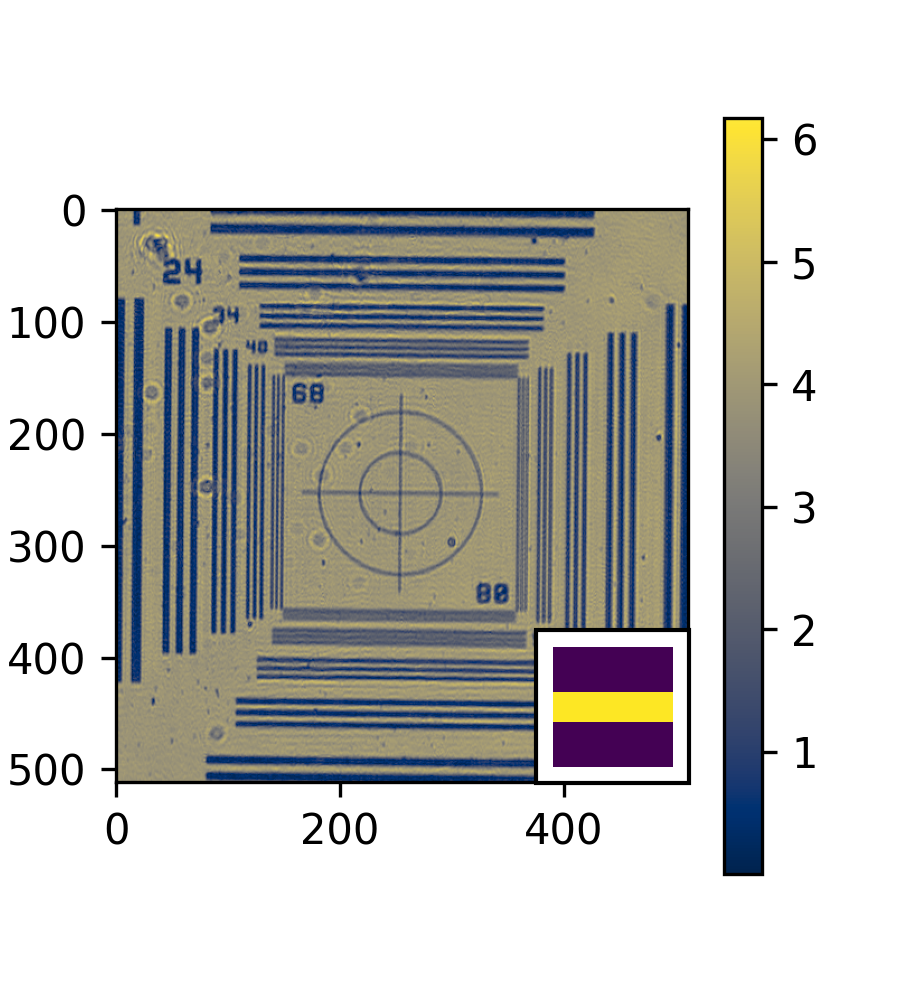}
        \caption{}
        \label{fig:512band}
    \end{subfigure}
    \vfill
    \begin{subfigure}[b]{0.49\columnwidth}
        \centering
        \includegraphics[width=\textwidth]{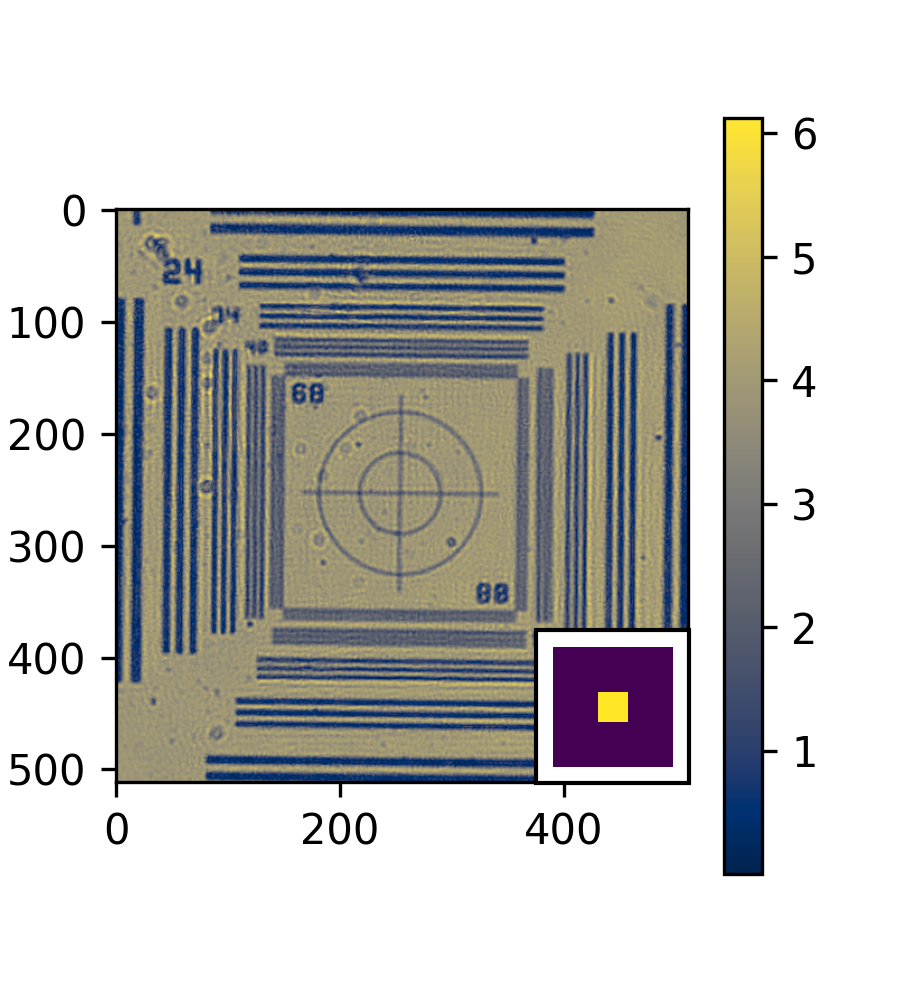}
        \caption{}
        \label{fig:512centre}
    \end{subfigure}
    \hfill
    \begin{subfigure}[b]{0.49\columnwidth}
        \centering
        \includegraphics[width=\textwidth]{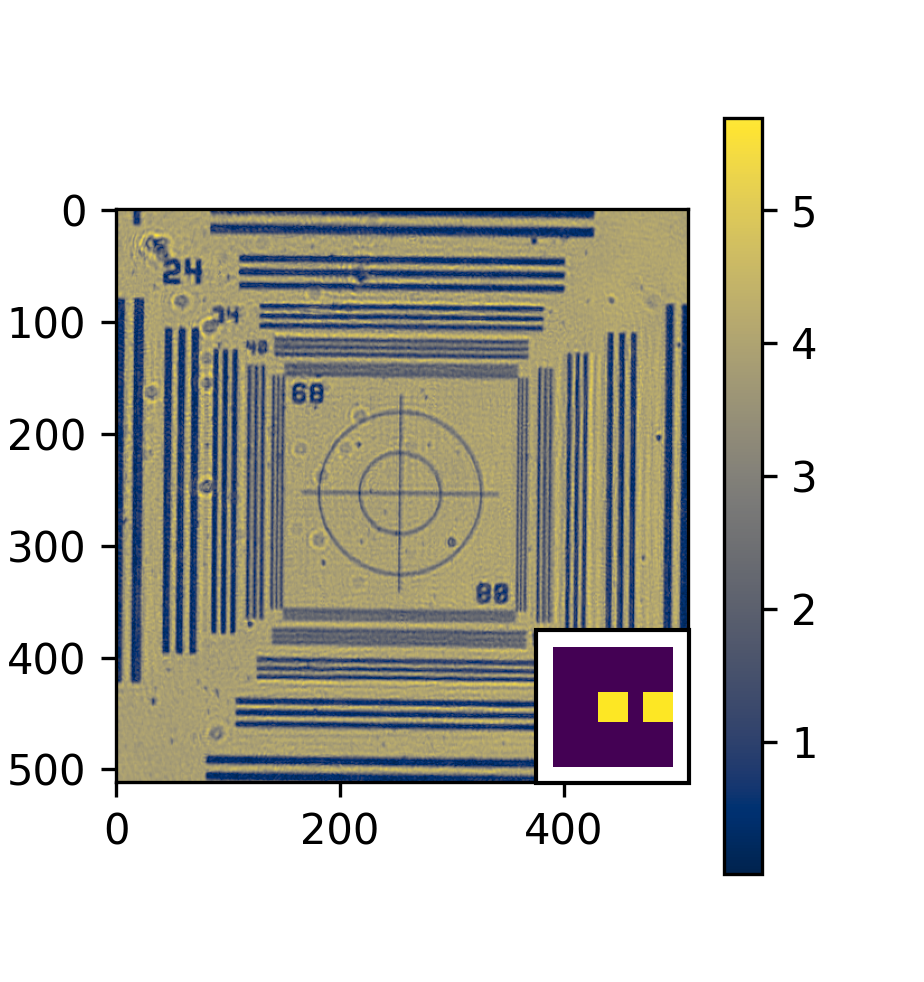}
        \caption{}
        \label{fig:512centre512}
    \end{subfigure}
    \caption{ \textbf{(a)} A reconstruction using data from the full-frame of the camera sensor. \textbf{(b)} A reconstruction using data from the full-frame width and a quarter of the full-frame height. \textbf{(c)} A reconstruction using a $512 \times 512$ sensor at the centre of the camera frame. Equivalent to the reconstruction using sensor $a$ in the main document and \textbf{(d)} a reconstruction using a $512 \times 512$ sensors at the centre and edge of the camera frame. Equivalent to using sensors $a$ and $b$.}\label{fig:mask1}
\end{figure}

Fig. \ref{fig:mask1} shows four reconstructions with different sensor configurations. As aid to understanding the effects of different configurations, Fig. \ref{fig:512centre} and \ref{fig:512centre512} show reconstructions equivalent to the results obtained by sensor fusion. 

A reconstruction using the full-frame of our detector is shown in Fig. \ref{fig:fullframe}. This reconstruction shows more detail than the reconstruction analogous to our reconstruction of sensor $a$ and sensor $b$. This is to be expected, as its signal contains spatial frequencies not present in the other reconstructions, namely the signal with spatial frequencies that contain information about the horizontal lines (vertical resolution). We note that taking into account the full-frame sensor data does not appreciably change the resolution on the set of finest vertical lines.

Fig. \ref{fig:512band} shows a reconstruction using a band with the full-frame width and a quarter of the full-frame height. Its resolution is an-isotropic, similar to the reconstruction using the two fused sensors. It has signal containing the spatial frequencies to reconstruct the vertical lines (horizontal resolution), but not the signal needed to reconstruct the horizontal lines (vertical resolution).

\begin{figure}[H]
    \centering
    \begin{subfigure}[b]{0.49\columnwidth}
        \centering
        \includegraphics[width=\textwidth]{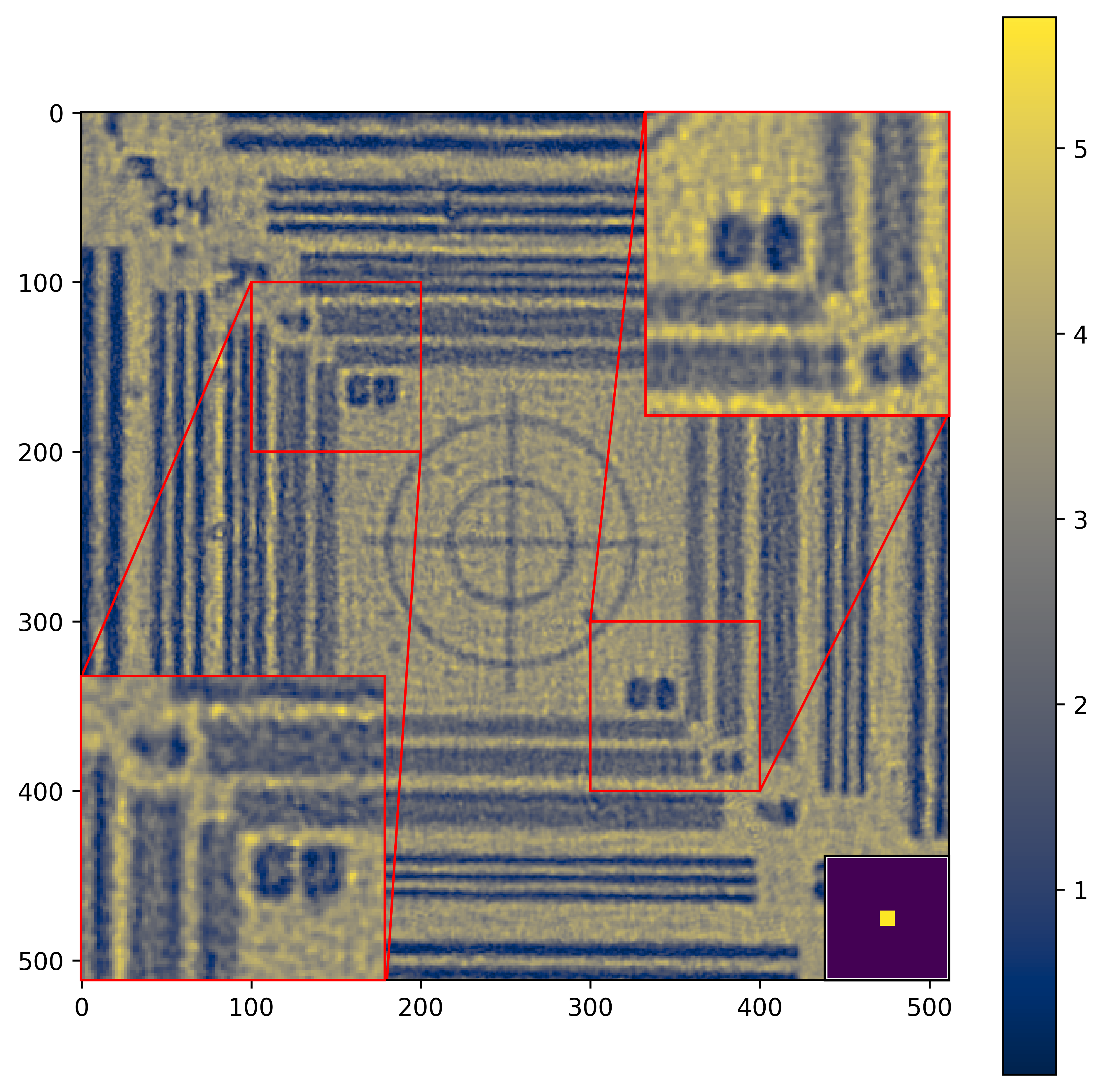}
        \caption{}
        \label{fig:256}
    \end{subfigure}
    \hfill
    \begin{subfigure}[b]{0.49\columnwidth}
        \centering
        \includegraphics[width=\textwidth]{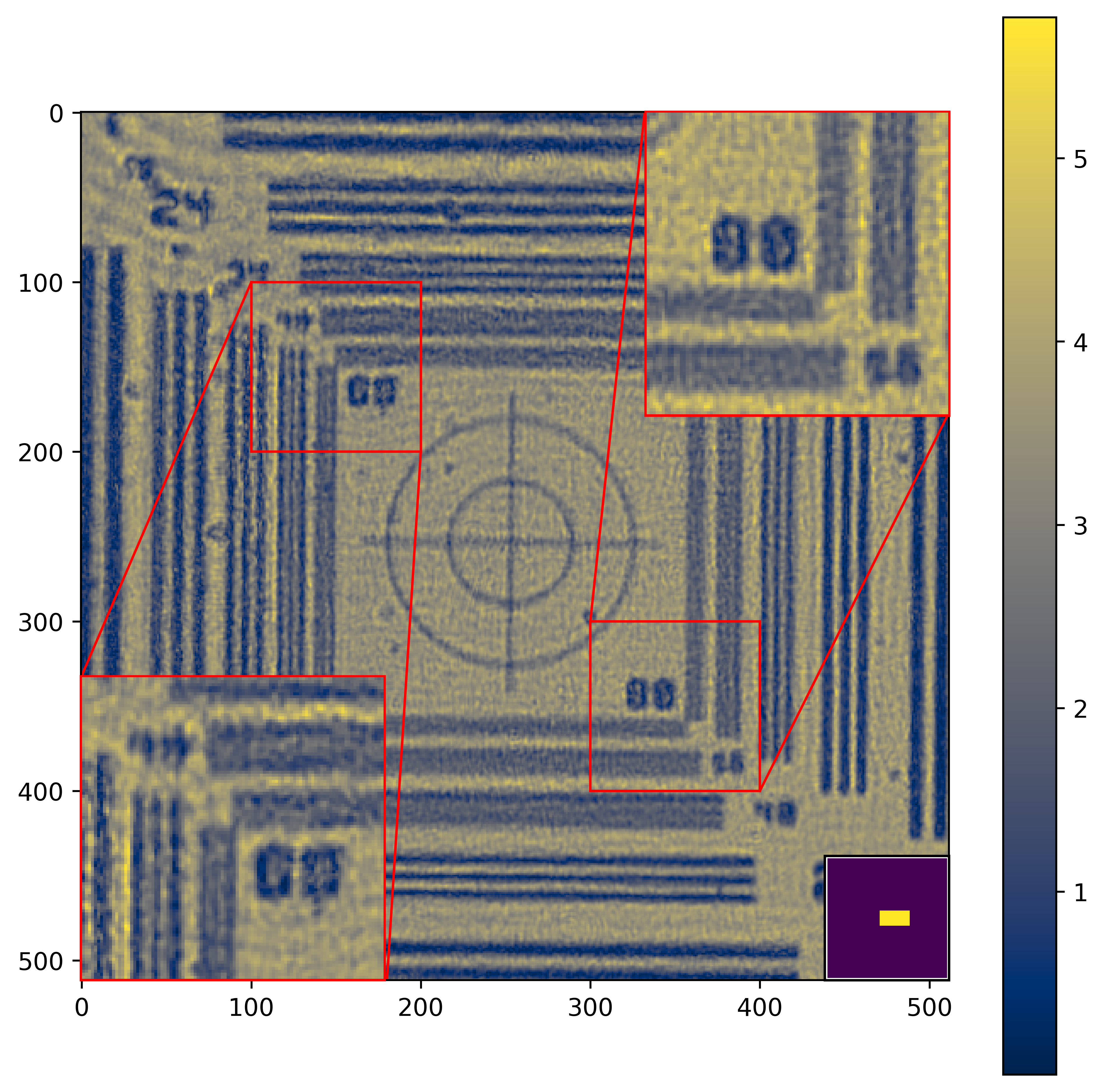}
        \caption{}
        \label{fig:256A}
    \end{subfigure}
    \vfill
    \begin{subfigure}[b]{0.49\columnwidth}
        \centering
        \includegraphics[width=\textwidth]{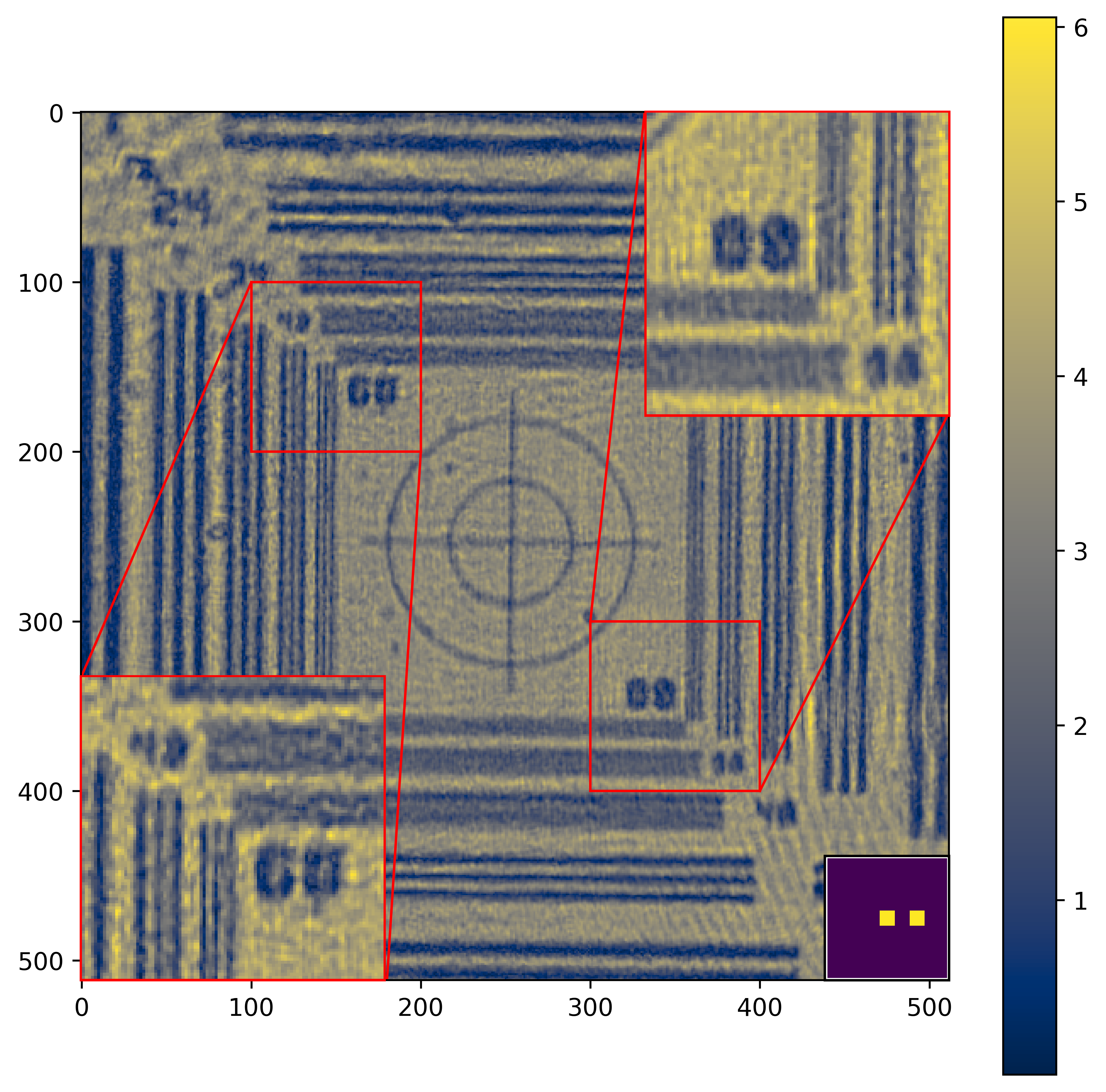}
        \caption{}
        \label{fig:256B}
    \end{subfigure}
    \hfill
    \begin{subfigure}[b]{0.49\columnwidth}
        \centering
        \includegraphics[width=\textwidth]{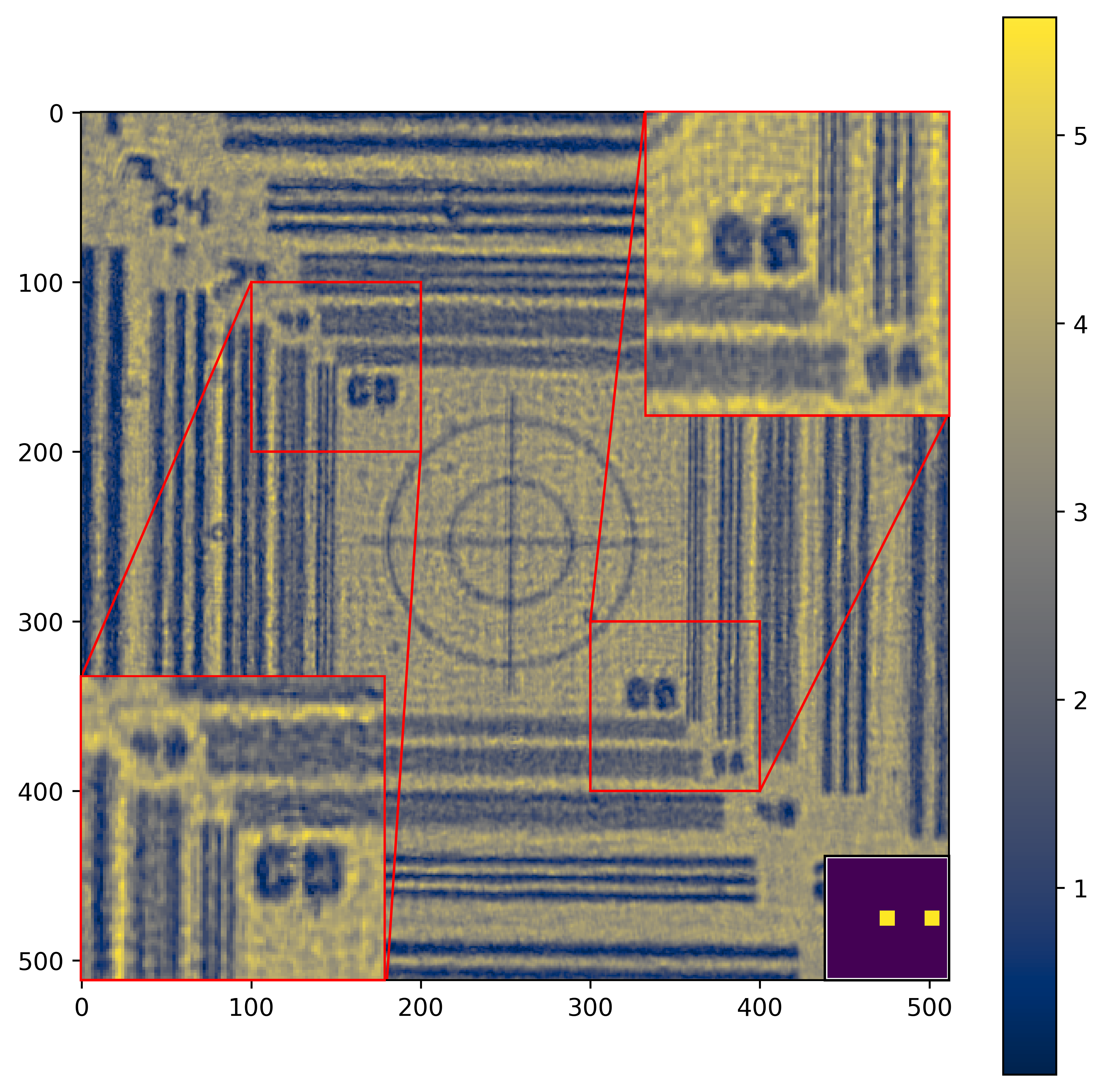}
        \caption{}
        \label{fig:256C}
    \end{subfigure}
    \caption{ \textbf{(a)} A reconstruction using a $256 \times 256$ sensor at the centre of the camera frame. \textbf{(b)} A reconstruction using a $256 \times 256$ sensor at the centre of the camera frame and one sensor horizontally shifted by \SI{883}{\micro\metre}. \textbf{(c)} A reconstruction using a $256 \times 256$ sensor at the centre of the camera frame and one sensor horizontally shifted by \SI{1766}{\micro\metre}. \textbf{(d)} A reconstruction using a $256 \times 256$ sensor at the centre of the camera frame and one sensor horizontally shifted by \SI{2649}{\micro\metre}.}\label{fig:mask2}
\end{figure}

Finer structures in the sample diffract light at higher angles. The detection of this signal is necessary to reconstruct these finer structures faithfully. However, a large gap between sensors means that certain spatial frequencies will not be present in the detected signal.

Fig. \ref{fig:mask2} shows four reconstructions using $256 \times 256$ sensors. Fig. \ref{fig:256} has one sensor centred, and figures \ref{fig:256A}, \ref{fig:256B} and \ref{fig:256C} have two sensors each with the second sensor being horizontally shifted by 883, 1766, 2649 \si{\micro\metre} respectively. Figures \ref{fig:256A} and \ref{fig:256B} both show a resolved, vertical pattern for 48\,lines/\si{\milli\metre}, whereas figures \ref{fig:256} and \ref{fig:256C} do not. The latter both do not have a sensor which captures the signal with spatial frequencies required to reconstruct these line patterns. However, Fig. \ref{fig:256C} shows a resolved, vertical pattern for 68 and 80\,lines/\si{\milli\metre}, as well as 40\,lines/\si{\milli\metre} and larger structures. As in the main document, the horizontal line patterns do not improve in any of these reconstructions. However, the reconstruction algorithm can be modified to work with more than two sensors quite simply.

\end{document}